\documentstyle[12pt,a4wide]{article}
\let\ssection=\section
\renewcommand{\section}{\setcounter{equation}{0}\ssection}

\newcommand{\be}{\begin{enumerate}}
\newcommand{\ee}{\end{enumerate}}
\newcommand{\bi}{\begin{itemize}}
\newcommand{\ei}{\end{itemize}}
\newcommand{\beq}{\begin{equation}}
\newcommand{\eeq}{\end{equation}}
\newcommand{\beqa}{\begin{eqnarray}}
\newcommand{\eeqa}{\end{eqnarray}}

\renewcommand{\a}{\alpha}                   
\renewcommand{\b}{\beta}                    
\renewcommand{\d}{\delta}               
\newcommand{\g}{\gamma}

\newcommand{\ket}[1]{| #1 \rangle}

\newcommand{\tr}{{\rm\, tr}}

\newcommand\mathC{\mkern1mu\raise2.2pt\hbox{$\scriptscriptstyle|$}
                {\mkern-7mu\rm C}}
\newcommand{\mathR}{{\rm I\! R}}                

\newcommand{\C}{{\tilde{C}}}
\newcommand{\D}{{\cal D}}
\renewcommand{\P}{{\cal P}}

\newcommand{\UP}{{\cal UP}}
\renewcommand{\H}{{\cal H}}
\newcommand{\V}{{\cal V}}

\begin{document}

\begin{titlepage}
\hspace{10truecm} Imperial/TP/95--96/49\\
\hspace*{11,3truecm} gr-qc/9607050
\vspace*{0,5cm}
\begin{center}
   {\Large\bf Symmetries of Decoherence Functionals} 
\end{center}
\medskip
\begin{center}
                S.~Schreckenberg\footnote{email: stschr@ic.ac.uk}\\[0.5cm]
                Blackett Laboratory\\
                Imperial College\\
                South Kensington\\
                London SW7 2BZ
\end{center}
\medskip
\begin{center} May 1996 \\[0,5cm] {\em Submitted to JMP} \end{center}
\vspace{4cm}
\begin{abstract}
The basic ingredients of the `consistent histories' approach to
quantum theory are a space $\UP$ of `history propositions' and a space
$\D$ of `decoherence functionals'. In this article we consider such
history quantum 
theories in the case where  $\UP$ is given by the set of projectors $\P(\V)$ 
on some Hilbert space $\V$. Using an analogue of Wigner's Theorem in the
context of history quantum theories proven earlier, we
develop the  notion of a  
`symmetry of a decoherence functional' and  prove that all such
symmetries form a group which we  call `the symmetry group of a
decoherence functional'. We calculate---for the case of history quantum
mechanics---some of these symmetries explicitly and relate them to
some discussions that have appeared previously.\\ 


\end{abstract}
\vfill
\end{titlepage}


\section{Introduction}

The decoherent histories approach to quantum theory has received much
attention over the last years. One of its main features is that  
the notion of a history proposition as an entity is build into the
framework for 
such a theory from the very start. This amounts to an entirely new
approach to quantum theory as has been discussed by various authors 
\cite{Gri84,Omn88a,GH90a,I94a}.  \\

In \cite{I94a} Isham argued---by looking at
standard quantum mechanics from the point of view of the history
programme---that the mathematical 
structure of such theories is best described by separating the
ingredients into a space of history propositions, $\UP$, and a space of
decoherence functionals, $\D$, both of which can be specified  with the
aid of certain algebraic properties. In the case of finite--dimensional
quantum mechanics when investigated at a finite sequence of $n$
time--ordered, 
but otherwise arbitrary, time points $t_1< t_2 <\cdots < t_n$, 
these spaces are given by $\UP=\P(\V_n)$ the set of {\em Schr\"odinger
picture} projectors on the
$n$--fold tensor product space
$\V_n:=\H_{t_1}\otimes\H_{t_2}\otimes\cdots\otimes\H_{t_n}$ of the
single time Hilbert spaces $\H_{t_i}$. The `histories', that is
time--ordered sequences of Schr\"odinger picture projection operators
$(\a_{t_1},\a_{t_2},\ldots ,\a_{t_n})$ commonly used in the 
formalism of decoherent histories, are uniquely associated with a
subset of $\P(V_n)$, that is given by homogeneous  
projection operators of the form 
$\a_h:=\a_{t_1}\otimes\a_{t_2}\otimes\cdots\otimes\a_{t_n}\in\P(\V_n)$. 
The classification theorem for
decoherence functionals $d\in\D$  \cite{ILS94c} shows
that---for an arbitrary finite--dimensional Hilbert space
$\V$---complex--valued, bounded  decoherence functionals $d:
\P(\V)\times\P(\V)  
\rightarrow \mathC$, which have to fulfill the requirements
of
\begin{eqnarray}\label{r}
&\circ & {\em Hermiticity\/}: d(\a,\b)=d(\b,\a)^\ast\quad\forall
          \a,\b\in \P(\V)\\
&\circ & {\em Positivity\/}: d(\a,\a)\ge0\quad\forall \a\in \P(\V)\nonumber\\
&\circ & {\em Additivity\/}: d(\a\oplus\b,\g)=d(\a,\g)+d(\b,\g)\nonumber\\
&\circ & {\em Normalisation\/}: d(1,1)=1,\nonumber
\end{eqnarray}
are in one--to--one correspondence with certain operators
$X=X_1+iX_2 \in {\cal X}_\D$
on $\V\otimes\V$ according to the 
rule:
\beq\label{tr}
d(\a,\b)=\tr_{\V\otimes\V}(\a\otimes\b X)
\eeq
with the restriction that
\beqa
&a)& X^\dagger=MXM
\quad\mbox{with}\quad\!\!\!M(\ket{v}\otimes\ket{w}):=
\ket{w}\otimes\ket{v},\quad\!\!\!
\forall \ket{v},\ket{w}\in\V.\label{hr}\\
&b)& \tr_{\V\otimes\V}(\a\otimes\a X_1)\ge0 \label{b}\nonumber\\ 
&c)& \tr_{\V\otimes\V}(X_1)=1.\nonumber
\eeqa
${\cal X}_\D$ denotes the set of all such operators $X_d$. In
particular this holds true if $\V=\V_n$.  
For standard quantum mechanics---when looked at from the 
perspective of the
history programme---the decoherence functional is associated with an
operator
\beq
X_{(H,\rho_{t_0},\rho_{t_f})}=\frac{1}{\tr_\H
(\rho_{t_0}\rho_{t_f}(t_f))} \tilde{X}_{(H,\rho_{t_0},\rho_{t_f})}
\eeq
on $\V\otimes\V$. For
homogeneous propositions about histories
$\a_h:=\a_{t_1}\otimes\a_{t_2}\otimes\cdots\otimes\a_{t_n}$ the value
of $d_{(H,\rho_{t_0},\rho_{t_f})}(\a_h,\b_h)$ is evaluated to be 
\beqa\label{sd}
   d_{(H,\rho_{t_0},\rho_{t_f})}(\a_h,\b_h)&=& 
\frac{1}{\tr_\H(\rho_{t_0}\rho_{t_f}(t_f))}\tr_{\V\otimes\V}
(\a_h\otimes\b_h \tilde{X}_{(H,\rho_{t_0},\rho_{t_f})})\\
&=& \frac{1}{\tr_\H(\rho_{t_0}\rho_{t_f}(t_f))}
\tr_\H(\C_{\a_h}^\dagger\rho_{t_0} \C_{\b_h}\rho_{t_f}(t_f)),\nonumber 
\eeqa
where the  `class' operator $\C_{\a_h}$ \cite{GH90a} is defined to be 
\beq
\C_{\a_h}:=\a_{t_1}(t_1)\a_{t_2}(t_2)\cdots\a_{t_n}(t_n)
\eeq
with
$\{\a_{t_i}(t_i):=
e^{\frac{i}{\hbar}H(t_i-t_0)}\a_{t_i}e^{-\frac{i}{\hbar}H(t_i-t_0)}\}$
being the associated Heisenberg picture operators. It is this
expression from which the operator 
$X_{(H,\rho_{t_0},\rho_{t_f})}$ originally had been derived. The operator
$\tilde{X}_{(H,\rho_{t_0},\rho_{t_f})}$---defined on
$\V\otimes\V$---associated with the decoherence functional 
$d_{(H,\rho_{t_0},\rho_{t_f})}\in\D$ is given by
\cite{SS96a} 
\begin{eqnarray}\label{dhqm}
\tilde{X}_{(H,\rho_{t_0},\rho_{t_f})} &=&
\big[U(t_1,t_0)^\dagger
\rho_{t_0}U(t_1,t_0)\otimes U(t_2,t_1)^\dagger\otimes\cdots\otimes
U(t_{n},t_{n-1})^\dagger\big]\nonumber\\
&\otimes & 
\big[U(t_2,t_1)\otimes U(t_3,t_2)\otimes\cdots\otimes U(t_{n},t_{n-1})\otimes
 U(t_f,t_n)\rho_{t_f} U(t_f,t_n)^\dagger\big]\nonumber\\[2mm]
&\times & \big( R_{(n)}\otimes 1_{t_1}\otimes 1_{t_2}\otimes\cdots\otimes
1_{t_n}\big)\\
&\times &S_{(2n)} \nonumber\\
&\times &\big( R_{(n)}\otimes 1_{t_1}\otimes
1_{t_2}\otimes\cdots\otimes 1_{t_n}\big),  \nonumber
\end{eqnarray}
where the last three lines involve universal operators $R_{(n)}, S_{(2n)}$
that arise by rewriting products of operators---as they appear in
equation (\ref{sd})---in terms of
tensor--products via the use of the mathematical identity \cite{ILS94c} 
\beq \label{mi}
\tr_\H (A_1A_2\cdots A_m)=\tr_{\otimes\H^m}(A_1\otimes\cdots\otimes A_m
S),  
\eeq
where $A_i$ denote operators on the Hilbert space $\H$ and $S$
represents a certain universal operator on $\otimes\H^m$.
Therefore, the operators $R_{(n)}, S_{(2n)}$ are system
independent. Thus, this operator
$\tilde{X}_{(H,\rho_{t_0},\rho_{t_f})}$ contains essentialy only the initial
and final density 
operators $\rho_{t_0},\rho_{t_f}$ and the time--evolution operator
$U(t_{i},t_{i-1})$. For a more detailed account of the formalism used
here see \cite{I94a,SS96a}.\\

One important result of this separation into a space of history
propositions $\UP$  and decoherence functionals $\D$ is that the
decoherence functional $d\in\D$ can be thought of as the `dynamical' 
content of
a history quantum theory. In the case of standard 
quantum mechanics looked at from the point of view of the history
programme one sees explicitly that $d\_{(H,\rho_{t_0},\rho_{t_f})}in\D$ 
carries the knowledge of the initial and  final conditions as well
as of the Hamiltonian. History propositions $\a\in\P(\V)$ are given 
by Schr\"odinger--picture projection operators and provide thus only the
`kinematical' input of the quantum theory: their properties specify
the Hilbert space $\V$. \\

In standard quantum
mechanics symmetries of a physical system
represented by a Hamiltonian operator $H$ on a Hilbert space $\H$ are
often described in terms 
of operators $A$ on $\H$ that commute with $H$. In order to understand 
how
this comes about it is crucial to dinstinguish in standard quantum
mechanics between the notions of a {\em symmetry}, a
{\em physical symmetry} and a {\em symmetry of an operator}.\\

On a complex Hilbert space $\H$ symmetries---as
defined by Wigner---are represented by unitary
or anti-unitary operators $U$ on $\cal H$ that are characterized by
the property of 
leaving the modulus of the inner product of  any pair of two vectors
$\ket{v}, \ket{w}$ invariant, that is 
\beq
|\langle v, w\rangle |^2 = |\langle Uv, Uw\rangle |^2 .
\eeq
{\em Wigner's theorem} asserts that a {\em physical symmetry}---that
is an affine one--to--one map on the space of rays in a Hilbert space, that
preserves orthogonality between the rays---are in one--to--one
correspondence with symmetries and can thus be implemented by a
unitary or anti--unitary operator on $\H$.\\
 
Given an operator $A$ on $\cal H$ the  {\em symmetries of the operator}
$A$ are then defined to be all those symmetries $U^A$, that
commute with $A$, that is $[U^A,A]=0$. As a result, it holds that  
\beq\label{t}
|\langle v|A|w\rangle | = |\langle U^Av|A|U^Aw\rangle | \qquad\forall
v,w\in\cal H ,
\eeq

In the
case of quantum mechanics, the Hilbert space $\H$ is given by the
Hilbert space $\H_t$ at a single time--point $t\in\mathR$. In
case that $A$ is given by the unitary evolution 
operator $U(t,t_0)=e^{-iH(t-t_0)}$ of a quantum mechanical system,
the symmetries of this operator are given by all unitary operators
$e^{iK}$ on $\H_t$ which commute with $U(t,t_0)$. On the level of the
(anti--) hermitian operators $K$ and $H$ one requires that $[K,H]=0$.\\

We would like to understand  how such concepts  find their place
in a theory that 
places its emphasis on `history propositions' and `decoherence
functionals'. 
 Recall that the two main ingredients of a history quantum theory are
the space of 
history propositions $\UP$ that---in the case we are considering---is 
given by the set of projectors
$P(\V)$ on a finite--dimensional Hilbert space $\V$,   and the space of
decoherence 
functionals $\D$. 
Decoherence functionals $d\in\D$ are associated with operators $X_d\in
{\cal X}_\D$
on $\V\otimes\V$ and both are intertwined through  the expression
$d(\a,\b)=\tr_{\V\otimes\V}(\a\otimes\b X_d)$.\\

In a companion paper \cite{SS96a} we proved an analogue of Wigner's
theorem for 
history quantum theories.  We defined the notion of a `homogeneous
symmetry' (HS) and of a `physical
symmetry of a history quantum theory' (PSHQT) and showed 
that PSHQT are in one--to--one correspondence with HS. Therefore,
every PSHQT 
can be 
induced by a unitary operator $\hat{U}\otimes\hat{U}$ on $\V\otimes\V$ as 
follows:
\begin{eqnarray}
&\UP\times\UP&: \qquad\a\otimes\b\mapsto
\tilde{\a}\otimes\tilde{\b}:=
\hat{U}\a\hat{U}^\dagger\otimes\hat{U}\b\hat{U}^\dagger 
\qquad\forall \a\in\P(\V)\\
&{\cal X}_\D&    : \qquad\quad X_d\mapsto 
   X_{\tilde{d}}:=(\hat{U}\otimes\hat{U})X_d(\hat{U}^\dagger
\otimes\hat{U}^\dagger) .\nonumber 
\end{eqnarray}
As a consequence of this transformation the invariance  
\beq
d(\a,\b)=\tilde{d}(\tilde{\a},\tilde{\b})
\eeq
for all $d\in\D$ and all $\a ,\b\in P(\V)$ follows by the property of
the trace, see equation (\ref{tr}).  Thus, a  PSHQT possesses the 
properties of
(i) mapping $\UP\times\UP $ into itself {\em and} (ii) mapping ${\cal
X}_\D$ into itself {\em and} (iii) leaving the value $d(\a,\b)$
invariant for all $\a,\b\in\P(\V)$ and $d\in\D$. Two history quantum
theories that are related by a PSHQT are called `physically
equivalent'; this definition turns out to be compatible with the one
introduced in \cite{GH94}. \\ 
 
In 
section \ref{SD} we use this analogue of Wigner's theorem to
introduce  the  notion of `symmetries 
of a decoherence 
functional' in analogy to the definition of a symmetry of an operator
in standard quantum mechanics. These elements are shown to form a 
group which we call `the
symmetry group of a decoherence functional'. 
Furthermore, it is shown---by calculating for history quantum
mechanics some of these symmetries explicitly---how this definition
seems to capture the mathematical essence of some related discussions
that have  
appeared in the literature \cite{HLM94,DK94b}. In the closing section
\ref{Clo} we
mention some ways one could try to proceed in order to find a
satisfactory {\em physical}
interpretation of the symmetries considered in this article.

\section{Symmetries of Decoherence Functionals}\label{SD}

\subsection{Definition}

Every history quantum theory is determined by the choice of a
particular decoherence functional $d\in\D$ that is kept fixed in course of the
investigation. We are therefore led to the following definition of
symmetries of decoherence functionals:\\

\noindent
{\em \bf Definition} For a fixed decoherence functional $d\in\D$, the
{\em symmetries of d} are determined by those unitary or anti--unitary
transformations 
$\hat{U}$ on $\V$, such that 
\beq\label{ir}
d(\a,\b)=d(\hat{U}\a\hat{U}^\dagger,\hat{U}\b\hat{U}^\dagger)\quad\forall
\a,\b \in \P(\V). 
\eeq  
The following proposition shows that symmetries of a decoherence
functional possess a convenient characterization in terms of commuting
operators.\\

\noindent
{\bf Proposition} The set $S_d$ of symmetries of a decoherence
functional $d\in\D$ is given by 
\beq\label{us}
S_d:=\{ \hat{U}\otimes\hat{U}\in Aut(\V\otimes\V) :
[X_d,\hat{U}\otimes\hat{U}]=0 \}. 
\eeq
Thus, for every finite--dimensional Hilbert space $\V , X_d\in
{\cal X}_\D$ it holds that :
\beq\label{eq}
[X_d,\hat{U}\otimes\hat{U}]=0 \Leftrightarrow
\tr_{\V\otimes\V}(\a\otimes\b [X_d - 
\hat{U}^\dagger\otimes\hat{U}^\dagger X_d\hat{U}\otimes\hat{U}])=0
\quad\forall
\a,\b\in \P(\V). 
\eeq

\noindent
{\bf Proof}\\
Since $X_d=X_1+iX_2$ we have to show commutativity for the real and
imaginary part separately.
In \cite{SS96a} it was shown that every $X_{1}$ possesses
an expansion in terms of operators of the form  $P_{\ket{e_i}}\otimes
P_{\ket{e_j}} +  P_{\ket{e_j}}\otimes P_{\ket{e_i}}$ for some
orthonormal basis $\{\ket{e_i}\}$ of $\V$, and that every $X_2$ can be
expanded in terms of operators $P_{\ket{b_i}}\otimes
P_{\ket{b_j}} -  P_{\ket{b_j}}\otimes P_{\ket{b_i}}$ for some---in
general---different 
orthonormal basis $\{\ket{b_i}\}$ of $\V$.\\

Thus, for the real part we have to show that 
\beqa
& &[P_{\ket{e_i}}\otimes P_{\ket{e_j}} +  P_{\ket{e_j}}\otimes
P_{\ket{e_i}},\hat{U}\otimes\hat{U}]=0\\ 
&\Leftrightarrow& \tr_{\V\otimes\V}(\a\otimes\b[P_{\ket{e_i}}\otimes
P_{\ket{e_j}} + P_{\ket{e_j}}\otimes P_{\ket{e_i}} -
(P_{\ket{e_i}}^{\hat{U}}\otimes 
P_{\ket{e_j}}^{\hat{U}}- P_{\ket{e_j}}^{\hat{U}}\otimes
P_{\ket{e_i}}^{\hat{U}})]) = 0 \quad \forall \a,\b \in \P(\V), \nonumber 
\eeqa 
whereas for the imaginary part we must show that 
\beqa
& &[P_{\ket{b_i}}\otimes P_{\ket{b_j}} -  P_{\ket{b_j}}\otimes
P_{\ket{b_i}},\hat{U}\otimes\hat{U}]=0\\ 
&\Leftrightarrow& \tr_{\V\otimes\V}(\a\otimes\b[P_{\ket{b_i}}\otimes
P_{\ket{b_j}} - P_{\ket{b_j}}\otimes P_{\ket{b_i}} -
(P_{\ket{b_i}}^{\hat{U}}\otimes 
P_{\ket{b_j}}^{\hat{U}}- P_{\ket{b_j}}^{\hat{U}}\otimes
P_{\ket{b_i}}^{\hat{U}})]) = 0 \quad \forall \a,\b \in \P(\V). \nonumber 
\eeqa 
We first consider the real part of $d\in\D$.
By assumption, it is true that 
\beqa
& & \tr_\V(\a P_{\ket{e_i}})\tr_\V(\b P_{\ket{e_j}})+\tr_\V(\a
P_{\ket{e_j}})\tr_\V(\b P_{\ket{e_i}})\\
&=& \tr_\V(\a P_{\ket{e_i}}^{\hat{U}})\tr_\V(\b P_{\ket{e_j}}^{\hat{U}})+
\tr_\V(\a P_{\ket{e_j}}^{\hat{U}})\tr_\V(\b P_{\ket{e_i}}^{\hat{U}})\quad
\forall \a,\b \in \P(\V)\nonumber .
\eeqa
Since this has to hold for {\em all} $\a,\b\in \P(\V)$ we choose now:
\bi
\item $\a =P_{\ket{e_i}}$ and $\b =P_{\ket{e_j}}$. Thus,  
\beq
1=\tr_\V(P_{\ket{e_i}}P_{\ket{e_i}}^{\hat{U}})\tr_\V(P_{\ket{e_j}}
P_{\ket{e_j}}^{\hat{U}})+ \tr_\V( P_{\ket{e_i}} P_{\ket{e_j}}^{\hat{U}})
\tr_\V(P_{\ket{e_j}} P_{\ket{e_i}}^{\hat{U}}).\label{rd1}
\eeq
\item $\a =P_{\ket{e_i}}=\b$. Thus
\beq
\tr_\V(P_{\ket{e_i}}P_{\ket{e_i}}^{\hat{U}})\tr_\V(P_{\ket{e_i}}
P_{\ket{e_j}}^{\hat{U}})=0.
\eeq
\item $\a =P_{\ket{e_j}}=\b$. Thus 
\beq
\tr_\V(P_{\ket{e_j}}P_{\ket{e_j}}^{\hat{U}})\tr_\V(P_{\ket{e_j}}
P_{\ket{e_i}}^{\hat{U}})=0.
\eeq
\ei
Assuming that $\tr_\V(P_{\ket{e_i}}P_{\ket{e_i}}^{\hat{U}})=0$ leads via
equation (\ref{rd1}) to the condition that 
\beq
1=\tr_\V( P_{\ket{e_i}}
P_{\ket{e_j}}^{\hat{U}})\tr_\V(P_{\ket{e_j}} P_{\ket{e_i}}^{\hat{U}}),
\eeq
that, since $\hat{U}$ is either unitary or anti--unitary,  can only be
fulfilled for $\hat{U}$ being a transposition. But, since this has to
hold for all orthogonal pairs $\ket{e_i},\ket{e_j}$, this contradicts
the unitarity or anti--unitarity of $\hat{U}$. Therefore, we 
conclude that $\tr_\V(P_{\ket{e_i}}P_{\ket{e_j}}^{\hat{U}})=0$. This
leads to the condition 
\beq\label{A1a}
1=\tr_\V(P_{\ket{e_i}}P_{\ket{e_i}}^{\hat{U}})\tr_\V(P_{\ket{e_j}}
P_{\ket{e_j}}^{\hat{U}}).
\eeq
Since $\hat{U}$ is a {\em unitary or anti--unitary} operator, it
follows that 
\beq
|\tr_\V(P_{\ket{e_i}}P_{\ket{e_i}}^{\hat{U}})|\le 1 \quad\forall
\ket{e_i}\in\V. 
\eeq
This shows that the equality (\ref{A1a}) holds if and only if 
$[P_{\ket{e_i}}\otimes P_{\ket{e_j}},\hat{U}\otimes\hat{U}]=0.$ This
concludes the proof for the real part.\\

For the imaginary part we are led to the condition 
\beqa
& & \tr_\V(\a P_{\ket{b_i}})\tr_\V(\b P_{\ket{b_j}})-\tr_\V(\a
P_{\ket{b_j}})\tr_\V(\b P_{\ket{b_i}})\\
&=& \tr_\V(\a P_{\ket{b_i}}^{\hat{U}})\tr_\V(\b 
P_{\ket{b_j}}^{\hat{U}})-\tr_\V(\a P_{\ket{e_b}}^{\hat{U}})
\tr_\V(\b P_{\ket{b_i}}^{\hat{U}})\quad
\forall \a,\b \in \P(\V)\nonumber .
\eeqa
Since this has to hold for {\em all} $\a,\b\in \P(\V)$ we choose now:
\bi
\item $\a =P_{\ket{b_i}}$ and $\b =P_{\ket{b_j}}$. Thus,  
\beq
1=\tr_\V(P_{\ket{b_i}}P_{\ket{b_i}}^{\hat{U}})\tr_\V(P_{\ket{b_j}}
P_{\ket{b_j}}^{\hat{U}})- \tr_\V( P_{\ket{b_i}} P_{\ket{b_j}}^{\hat{U}})
\tr_\V(P_{\ket{b_j}} P_{\ket{b_i}}^{\hat{U}}).\label{rd2}
\eeq
\ei
Since---by applying Gleason's theorem to the Hilbert space
$\V\otimes\V$---it holds that  
\beq
\tr_\V( P_{\ket{b_i}} P_{\ket{b_j}}^{\hat{U}})\tr_\V(P_{\ket{b_j}}
P_{\ket{b_i}}^{\hat{U}})=\tr_{\V\otimes\V}(P_{\ket{b_i}}\otimes
P_{\ket{b_j}}
[\hat{U}^\dagger\otimes\hat{U}^\dagger(P_{\ket{b_j}}\otimes P_{\ket{b_i}})
\hat{U}\otimes\hat{U}])\ge 0,
\eeq
it follows that 
\beq
\tr_\V(P_{\ket{b_i}}P_{\ket{b_i}}^{\hat{U}})\tr_\V(P_{\ket{b_j}}
P_{\ket{b_j}}^{\hat{U}})\ge 1\quad\forall i,j\in\{1,2,\ldots,{\dim\V}\}.
\eeq
But, since $\hat{U}$ is unitary or anti--unitary, the number $1$ is the
maximum from which we conclude that
$\tr_\V(P_{\ket{b_i}}P_{\ket{b_j}}^{\hat{U}})
\tr_\V(P_{\ket{b_j}}P_{\ket{b_i}}^{\hat{U}})=0$.
By the same reasoning as before this leads to $[P_{\ket{b_i}}\otimes
P_{\ket{b_j}},\hat{U}\otimes\hat{U}]=0$ which concludes the proof for
the imaginary part.\hfill $\Box$


\subsection{Discussion}

Since a symmetry of a decoherence functional is represented by a {\em
unitary} or {\em anti--unitary} operator $\hat{U}$, it preserves the
algebraic relations among history propositions
$\a\in\P(\V)$. For example, orthogonal elements $\a,\b\in\P(\V)$ are mapped
into elements $\a^\prime,\b^\prime$ that are also
orthogonal. Therefore, because it also preserves the value of the
decoherence functional on pairs of history propositions, consistent
sets of $d\in\D$ are mapped into new consistent sets to which 
the {\em same}
values of $d\in\D$ are associated. Recall \cite{S95a} that in the 
formalism used
here, {\em consistent sets of
history propositions with
respect to a particular $d\in\D$} correspond to certain
partitions of the unit operator on $\V$ into mutually orthogonal
projectors $\{\a_i\}_{i=1}^{m\le\dim\V}$ such that
\beq
d(\a_i,\a_j)=\d_{ij}d(\a_i,\a_i)\quad\forall i,j\in\{1,2,\ldots ,m\}.
\eeq   
The properties (\ref{r}) of $d\in\D$ ensure that the values $d(\a_i,\a_i)$
determine
a probability distribution on the boolean algebra generated by the
$\{\a_i\}_{i=1}^{m\le\dim\V}$. Thus, one can think of
symmetries of a decoherence functional as a way of generating new
consistent sets from given ones. A study of the symmetries of a
decoherence functional will therefore reduce the number of algebraic 
equations
representing the consistency requirements.\\

Since $X_d=X_1+iX_2$, equation (\ref{us}) is equivalent to the two
conditions 
\beq
[X_1,\hat{U}\otimes\hat{U}]=0 \quad\mbox{and}\quad
[X_2,\hat{U}\otimes\hat{U}]=0, 
\eeq
which express the invariance requirement for the real and imaginary
part of $X_d$ respectively. Since for each $X_d$ its  real and
imaginary part can be expanded with the aid of two, in general, {\em
different} bases of $\V$, we see that requiring the complex value of
$d\in\D$ to be invariant is a much stronger requirement than
invariance of the values for the real part $\Re d$ alone. It is
interesting to note that the following Corollary holds.\\

\noindent
{\bf Corollary} The following two requirements for the invariance of
the real part of the decoherence functionals are equivalent:
\beqa
\Re d(\a,\b) &=& \Re
d(\hat{U}\a\hat{U}^\dagger,\hat{U}\b\hat{U}^\dagger)\quad\forall \a,\b
\in \P(\V)\\
\Leftrightarrow\quad
d(\a,\a) &=& d(\hat{U}\a\hat{U}^\dagger,\hat{U}\a\hat{U}^\dagger)\quad\forall
\a\in \P(\V).\nonumber 
\eeqa  
This shows that requiring the invariance of the `diagonal part' 
$d(\a,\a)$ forces the entire real part, $\Re d(\a,\b)$, of $d\in\D$ to
be invariant under a symmetry transformation. \\

\noindent
{\bf Proof}\\
The proof is similar to the one presented for the proposition. We start
with the expression $d(\a,\a)$. Expand the
real part $X_1$ of $X_d$ in a terms of elementary decoherence functionals
$\frac{1}{2}(P_{\ket{e_i}}\otimes P_{\ket{e_j}}+ P_{\ket{e_j}}\otimes
P_{\ket{e_i}})$ with respect to a certain basis 
$\{\ket{e_i}\}$ of $\V$. Evaluation of  $d(\a,\a)$ for 
$\a=P_{\ket{e_i}}$, $\a=P_{\ket{e_j}}$ and
$\a=P_{\ket{e_i}}+P_{\ket{e_j}}$ leads to
$[X_1,\hat{U}\otimes\hat{U}]=0$ which equals the condition for the
invariance of the values of $\Re d(\a,\b)$ for all $\a,\b\in\P(\V)$.\hfill
$\Box$\\

A slightly more subtle observation is the following: Since we are
dealing with a unitary operator which preserves the algebraic
relations among history propositions, the property of a consistent set
of being a partition of unity is {\em always} preserved under this
mapping. To preserve the {\em value} of the decoherence functionals on
the elements of the consistent sets, only the `diagonal' values of
$d\in\D$, $d(\a,\a)$, have to remain invariant. This leads to the
requirement 
\beq\label{irA}
d(\a,\a)=d(\hat{U}\a\hat{U}^\dagger,\hat{U}\a\hat{U}^\dagger)\quad\forall
\a\in \P(\V), 
\eeq 
which, as we have seen,  is {\em not} equivalent to the vanishing of 
the commutator,
i.e.~$[X_d,\hat{U}\otimes\hat{U}]=0 $. It only leads to the weaker 
condition 
\beq
\tr_{\V\otimes\V}(\a\otimes\a
X_d)=\tr_{\V\otimes\V}(\a\otimes\a [\hat{U}^\dagger
\otimes\hat{U}^\dagger X_d
\hat{U}\otimes\hat{U}]), \quad \forall \a\in\P(\V).
\eeq
Therefore,  the sets of
transformations $\{\hat{U}\}$ obtained by enforcing  condition
(\ref{ir}) or (\ref{irA}), can, {\em a priori}, be different. It
suggests itself to call the transformations determined by (\ref{irA}) 
{\em weak symmetries of
a decoherence functional},
since one only   requires the commutator 
$[X_d,\hat{U}\otimes\hat{U}]$  to {\em vanish weakly}, meaning that
\beq
[X_d,\hat{U}\otimes\hat{U}]=\Delta_{\hat{U}},
\eeq
where $\Delta_{\hat{U}}$ is any operator on $\V\otimes\V$, such that 
\beq
\tr_{\V\otimes\V}(\a\otimes\a \Delta_{\hat{U}})=0, \quad \forall \a\in\P(\V).
\eeq
This ensures that condition (\ref{irA}) is met. We denote the set of
weak symmetries of  a decoherence functional by $S_d^w$.  
Every $\hat{U}\in S_d$ is also a weak symmetry, i.e.~$S_d\subset
S_d^w$,  but the converse is in
general not true. In view of the Corollary above, these weak
symmetries commute with the real part of $X_1$ of $X_d$ so that this
is a condition on the commutator $[X_2,\hat{U}\otimes\hat{U}]$ between
the imaginary part $X_2$ and the symmetry transformation.\\

Weak symmetries possess the property of mapping a consistent set
$\{\a_i\}$ ,
i.e.~a set for which the consistency conditions hold, into a partition of
unity $\{\a_i^\prime\}$, for which $d(\a_i^\prime,\a_i^\prime)$
defines a probability distribution. But, 
it has not been shown that this new set is also {\em consistent},
i.e.~obeys $d(\a_i^\prime,\a_j^\prime)=0$ for all $i\neq j$. Thus we
are led to the 
question of whether or not the consistency conditions $d(\a_i,\a_j)=0$
determine {\em all} 
partitions of unity $\{\a_i\}$ on which $d(\a_i,\a_i)$ defines a
probability distribution. Only then can we be sure that weak
symmetries map consistent sets into other consistent sets. But this is
certainly {\em not} the case. One
convinces oneself immediately that one should be
investigating this question for the consistency condition $\Re
d(\a_i,\a_j)=0$, since 
the vanishing of the imaginary part of 
the decoherence functional is unimportant in this context. If this
question can be answered in the affirmative, this would be a strong
argument in favour of using $\Re d(\a_i,\a_j)=0$ as consistency
conditions. In that case, the property of definig a probability
distribution on a partition of unity would be equivalent to the
consistency conditions $\Re d(\a_i,\a_j)=0$.   \\


\subsection{The structure of $S_d$}

The purpose of this section is to introduce the notion of a `symmetry
group of $d\in\D$'.\\

\noindent
{\bf Definition/Proposition} The set 
$S_d:=\{ \hat{U}\otimes\hat{U}\in Aut(\V\otimes\V) :
[X_d,\hat{U}\otimes\hat{U}]=0 \}$ of symmetries of a 
decoherence functional $d\in\D$ possesses the structure of a
group. This group is 
called {\em the symmetry group of d}.\\

\noindent
{\bf Proof}\\
The unit element is given by the unit operator ${\bf 1}\otimes{\bf1}$.
Multiplication is defined by
multiplication of operators. A calculation shows that
$[X_d,\hat{U}\hat{V}\otimes\hat{U}\hat{V}]=0$ so that
$\hat{U}\hat{V}\otimes\hat{U}\hat{V}\in S_d$ whenever
$\hat{U}\otimes\hat{U},\hat{V}\otimes\hat{V} \in S_d$.
The unique inverse $(\hat{U}\otimes\hat{U})^{-1}$ is given by
$\hat{U}^\dagger\otimes\hat{U}^\dagger$. We have to show that
\beq
\hat{U}\otimes\hat{U}\in S_d \quad\Longleftrightarrow
\quad \hat{U}^\dagger\otimes\hat{U}^\dagger\in S_d.
\eeq
This can be shown as follows.
\beqa\label{eqs}
[X_d,\hat{U}\otimes\hat{U}]=0 
&\Longleftrightarrow& X_d\hat{U}\otimes\hat{U} - 
\hat{U}\otimes\hat{U}X_d=0\\
&\Longleftrightarrow&
\hat{U}^\dagger\otimes\hat{U}^\dagger X_d^\dagger -
X_d^\dagger\hat{U}^\dagger\otimes\hat{U}^\dagger=0 \nonumber\\
&\Longleftrightarrow&  \hat{U}^\dagger\otimes\hat{U}^\dagger MX_dM -
MX_dM\hat{U}^\dagger\otimes\hat{U}^\dagger=0 \nonumber,
\eeqa
since $X_d^\dagger=MX_dM$. Using the properties  $MM=1\otimes 1$ and
$M(\hat{U}^\dagger\otimes\hat{U}^\dagger )M=
\hat{U}^\dagger\otimes\hat{U}^\dagger$
one concludes that (\ref{eqs}) is equivalent to 
\beq
[\hat{U}^\dagger\otimes\hat{U}^\dagger,X_d]=0.
\eeq
This concludes the proof. \hfill $\Box$\\

{\em Remark}: The inverse is {\em unique} since we are looking at
{\em unitary} operators $\hat{U}\otimes\hat{U}$ on $\V\otimes\V$, even 
though $\hat{U}$
might be unitary or anti--unitary.\\

It is interesting to ask for the structure of the set $S_d^w$ of weak
symmetries of d. This set will in general not be 
closed under the operations of multiplication and taking the inverse
as defined above. A quick calculation shows that 
\beq
[X_d,\hat{U}\hat{V}\otimes\hat{U}\hat{V}]=\Delta_{\hat{U}}
(\hat{V}\otimes\hat{V})+(\hat{U}\otimes\hat{U}) \Delta_{\hat{V}},
\eeq
so that the commutator will, in general, not vanish weakly. It would be
interesting to see whether or not  for these
transformations a kind of `Dirac--bracket' can be introduced along the
following lines:
\beq
[X_d,\hat{U}\otimes\hat{U}]^D:=[X_d,\hat{U}\otimes\hat{U}]-\Delta_{\hat{U}}.
\eeq
Thus, it follows that 
\beq
[X_d,\hat{U}\hat{V}\otimes\hat{U}\hat{V}]^D=
\Delta_{\hat{U}}(\hat{V}\otimes\hat{V})+(\hat{U}\otimes\hat{U})
\Delta_{\hat{V}}-\Delta_{\hat{U}\hat{V}}.
\eeq
Defining an operator $\Delta$ on the space of operators  by $\Delta
(\hat{U}\otimes\hat{U}):=\Delta_{\hat{U}}$, we see that to require 
\beq
\Delta_{\hat{U}}(\hat{V}\otimes\hat{V})+(\hat{U}\otimes\hat{U})
\Delta_{\hat{V}}-\Delta_{\hat{U}\hat{V}}
\eeq
to vanish weakly is certainly fulfilled if $\Delta$ possesses the
property of being a derivation on the space of
transformations $\hat{U}\otimes\hat{U}$. In this case, the set $S_d^w$
would be closed under multiplication; if $\hat{U}$ is an element of
$S_d^w$, its inverse will, in
general, not be an element of $S_d^w$. It seems that some insight
could be gained by such an analysis once a formulation of constraint
analysis in the histories formalism is achieved.

\subsection{History Quantum Mechanics}

We want to investigate, for ordinary non--relativistic quantum
mechanics when looked at from the perspective of the history programme, 
whether or not this definition of symmetries is of any value. Again,
history propositions $\a\in\UP$ are given by projectors $\a\in
\P(\V_n)=\P(\otimes_{i=1}^n\H_{t_i})$. The particular decoherence
functional is given by expression (\ref{dhqm}).\\ 

Can we find any symmetries $\hat{U}$ on $\V_n$ of this
$d_{(H,\rho_{t_0},\rho_{t_f})}$ in the sense specified above?

\bi
\item  Consider a transformation $Q_{t_i}\in B(\H_{t_i})$,
$\a_{t_i}\mapsto Q_{t_i}\a_{t_i}Q_{t_i}^\dagger$  
on all Hilbert spaces $\H_{t_i}$ associated with $n$ time--points
$(t_1<t_2<\cdots <t_n)$ ;
choose 
$\hat{U}:=Q_{t_1}\otimes Q_{t_2}\otimes\cdots\otimes Q_{t_n}$.
Then
$X_d^\prime\equiv\hat{U}^\dagger\otimes\hat{U}^\dagger
X_d\hat{U}\otimes\hat{U} 
$ contains the following terms: 
\begin{eqnarray}
& &Q_{t_1}^\dagger U(t_1,t_0)^\dagger\rho_{t_0}U(t_1,t_0)Q_{t_1}
\nonumber\\
& &Q_{t_i}^\dagger U(t_i,t_{i-1})^\dagger Q_{t_{i-1}}\nonumber\\
& &Q_{t_i}^\dagger U(t_i,t_{i-1}) Q_{t_{i-1}}\nonumber\\
& &Q_{t_n}^\dagger U(t_f,t_n)\rho_{t_f} U(t_f,t_n)^\dagger
Q_{t_n}\nonumber
\end{eqnarray}
The requirement that $[\hat{U}\otimes
\hat{U},X_{(H,\rho_{t_0},\rho_{t_f})}]=0$ is 
certainly fulfilled if we
 choose the same unitary $Q$ for all times $t_i$ and require it to
commute with the unitary evolution operator $U(t_i,t_{i-1})$, that is with the
Hamiltonian $H$, $[Q,H]=0$. Furthermore, we
notice  that $Q$ has to commute with the initial  and final  density
matrices, i.e.~$[Q,\rho_{t_0}]=0=[Q,\rho_{t_f}]$.
\ei
These examples of symmetries of the standard decoherence functional
are the easiest one can find and are exactly those one obtains by
transforming the triple $(H,\rho_{t_0},\rho_{t_f})$ by an appropriate,
fixed unitary 
transformation on $\H$. One can read them off the form
(\ref{sd}) of the
decoherence functional almost immediately \cite{HLM94,DK94b}. The
virtue of the derivation 
presented here is that it is associated with a well defined operator
$\hat{U}=Q\otimes Q\otimes\cdots\otimes Q$, $n$ times, on $\V$ which
fulfills the requirement of definition (\ref{us}). Thus, a symmetry of
a decoherence functional is not just determined by operators $Q$ that
commute with the Hamiltonian at a fixed `time--slice'; they take the
initial and final conditions into account and are more
to be regarded as `space-time' symmetries of the system, since we have
tied together the properties at different time--slices via the
tensor product operation.\\ 

The appearance of the commutator of $Q$ with the Hamiltonian $H$
signals that Noether's theorem enters the stage in a somewhat
disguised form. An explicit understanding of the precise relationship
would certainly further the understanding of symmetries of
decoherence functionals as defined here. I give a few remarks in the
closing section.


\section{Summary and Outlook}\label{Clo}

In this article we used the analogue of Wigner's theorem in the
context of history quantum theories \cite{SS96a} in order to define
the notion of a 
`symmetry of a decoherence functional'. 
We have seen that these
symmetries can be characterized transparently in terms of the vanishing 
of the
commutator 
$[X_d,\hat{U}\otimes\hat{U}]=0$ 
between the operator $X_d$ uniquely associated with $d\in\D$ and the
operator $\hat{U}\otimes\hat{U}$ associated with the symmetry
transformation. It has been shown that the set of symmetries of a
decoherence functional forms a group, called `the symmetry group of
$d\in\D$'. We calculated explicitly some symmetries for the case 
of history quantum mechanics which could be related to certain examples
discussed in \cite{HLM94,DK94b}.\\

Physical symmetries of history quantum theories and symmetries of
decoherence functionals  have now aquired a definite
mathematical interpretation as transformations on the space of history
propositions and on the space of decoherence functionals. But this 
does not do justice to the importance
that symmetries play in almost every physical theory. One  would like
to use the results presented here to gain a better insight into the
{\em physical} 
meaning of PSHQT. Usually, this means to obtain an interpretation of
such symmetries of history quantum theories or decoherence functionals
at a {\em classical} level; an interpretation which is then happily
taken over to the `quantized' theory. Thus, we are led to try to
obtain, for example, history quantum mechanics by a certain process of
`quantization' from a `classical history theory' for classical
mechanics. It turns out that history quantum mechanics can indeed be
obtained this way via the aid of a {\em history group}
\cite{IL95a}. Symmetries of history quantum theories and decoherence
functionals then correspond 
to certain transformations on a `space of histories'.  Details will
appear elsewhere.\\  

Even though we gained some insight into the properties of symmetries
of decoherence functionals I did not attempt in this paper to tackle
questions related to the meaning of `conserved
quantities' in this history formalism. This is a difficult issue since
in its conception history quantum theories are 
timeless. These problems have to some extent been
discussed in \cite{HLM94} but a satisfactory explanation remains to be
found. While searching for such an explanation an investigation of
properties of symmmetries of a decoherence functional---as proposed
in this article---might provide further clues to unravel its full
significance. 


\bigskip
\noindent
{\large\bf Acknowledgements}

\noindent
I would like to thank Professor Dr.~C.J.~Isham for useful discussions.


\end{document}